# Acoustothermal Effect: Mechanism and Quantification of the Heat Source

Pradipta Kr. Das[1], and Venkat R. Bhethanabotla[1,*]

[1]Department of Chemical, Biological, and Materials Engineering, University of South Florida, Tampa, FL 33620-5350, USA

[*]Contact author: bhethana@usf.edu

We examined theoretically, experimentally and numerically the origin of the acoustothermal effect using a standing surface acoustic wave actuated sessile water droplet system. Despite a wealth of experimental studies and a few recent theoretical explorations, a profound understanding of the acoustothermal mechanism remains elusive. This study bridges the existing knowledge gap by pinpointing the fundamental causes of acoustothermal heating. Theory broadly applicable to any acoustofluidic system at arbitrary Reynolds numbers going beyond the regular perturbation analysis is presented. Relevant parameters responsible for the phenomenon are identified and an exact closed form expression delineating the underlining mechanism is presented. Furthermore, an analogy between the acoustothermal effect and electromagnetic heating is drawn, thereby deepening understanding of the acoustothermal process.

*Introduction*-The interaction between acoustic waves and fluids spawns an oscillatory fluid flow concomitant with the wave's propagation and in addition to that, a mean flow known as acoustic streaming emerges over time. While the attenuation of acoustic waves gives rise to the hydrodynamic body force driving acoustic streaming [1–4], the non-linear coupling of the acoustic fields with the fluid flow also generates heat via acoustothermal effect [5,6]. The absorption of the acoustic energy in fluids has garnered significant attention over the past years due to the complex molecular interactions associated with the phenomena. The acoustic intensity ($\mathbf{I}$) inside the fluids shows an exponential decay, described by $\mathbf{I} = \mathbf{I}_0 e^{-\alpha|\mathbf{r}|}$ where $\mathbf{r} = |\mathbf{r}|\mathbf{e_r}$ is the position vector ($\mathbf{e_r}$ is the unit vector) and $\mathbf{I}_0$ is the intensity at the source (i.e. $\mathbf{r} = \mathbf{0}$) [7]. With the seminal theory provided by Stokes [8] in 1845, a substantial corpus of literature has traditionally attributed acoustic attenuation in fluids solely to shear viscosity leveraging translational relaxation of the molecular degrees of freedom. Later, researchers have modified the attenuation coefficient elucidating additional mechanisms of energy dissipation via vibrational, rotational, and molecular conformation relaxation, attributable to the bulk viscosity of the fluid [9].

The acoustothermal effect is strongly related to the absorption of the acoustic energy into the fluid and its conversion into internal energy. Even though viscous dissipation is known to be responsible for the generation of acoustothermal effect [10], a comprehensive understanding of the mechanism and the quantification of the heating remain elusive, as most of the available literatures are experimental in nature [11–15] and therefore, lack theoretical insights. This knowledge gap in acoustofluidics motivates our study, aiming to delve deeper into the underlying mechanisms of acoustothermal heating resulting in a closed-form expression quantifying the heat source.

In this letter, we investigate the acoustothermal effect both theoretically and experimentally by utilizing a sessile water droplet system actuated via surface acoustic wave (SAW) fields. Our principal findings are as follows: (i) The propagation of acoustic power within the fluid is exclusively characterized by the acoustic Poynting vector, analogous to the behavior of electromagnetic waves. (ii) The acoustic power flows coherently along the direction of the acoustic wave propagation. (iii) The acoustothermal heat generation rate is determined by the negative divergence of the Poynting vector, entailing both the rate of increase in acoustic energy density and the power absorbed through viscous damping. (iv) The time-averaged heat generation rate is solely defined by the time-averaged power loss spawned by viscous damping. While in this work, we employ counterpropagating surface acoustic waves (SAWs), in investigating the temperature rise of a sessile droplet, our acoustothermal theory generalizes immediately to any kind of acoustofluidic system. The fundamental insights into the acoustothermal effect are of paramount relevance to the broader community working in microfluidics and acoustics, given the bidirectional coupling with thermoacoustic streaming, which significantly alters the flow field or vice versa. The findings of the present study will not only bridge the knowledge gap in acoustofluidics but also pave the way for designing advanced and complex acoustofluidic systems.

*Experimental method* - The experiments were performed on a 128° Y-cut X-propagating lithium niobate (LiNbO$_3$) surface acoustic wave (SAW) device, comprising of 20 pairs of interdigital transducers (IDTs) etched onto each side, where the standing SAW was generated via the superposition of two waves of equal wavelength and amplitude traveling towards each other. The device is fabricated using a conventional SAW fabrication process involving photoresist coating, followed by ultraviolet exposure, mask lithography, aluminum deposition (100 nm thick), and finally, lift-off. Standing SAW with wavelength $\lambda = 120 \mu m$, was generated by exciting the IDT pairs with an AC signal keeping the spacing between two adjacent IDTs and the IDT width, both as $\frac{\lambda}{4} = 30\mu m$. The actuation frequency of the SAW device for all experiments was determined to be 32.016 MHz, corresponding to the minimum insertion loss; see sec. S1 in the Supplemental Material (SM) [16]. A sinusoidal signal was generated by a signal generator (Rohde & Schwarz, model SMA100A), amplified by a power amplifier (Mini-Circuits, model TIA-1000-1R8), and split to connect both IDTs via a signal splitter (Figure 1a). Ultrapure water droplets of 2 μl or 5 μl were placed on the delay path, and the droplet contact angle was measured as $\varphi_c =$ 100.63°±0.48°; see sec. S2 in the SM [16]. An infrared thermal camera (FLIR, model-T62101) was positioned above the sessile droplet to monitor the droplet's temperature over time during the experiments. The experiments began with a 15-second stabilization period without SAW, followed by a 90-second exposure to SAW to reach a steady-state temperature, and another 90 seconds post-SAW to return to ambient temperature. These experiments were conducted with and without droplets to isolate heating effects. The thermal camera used in these experiments, records at 30 frames per second, and the captured images were extracted and analyzed using MATLAB's image processing tools.

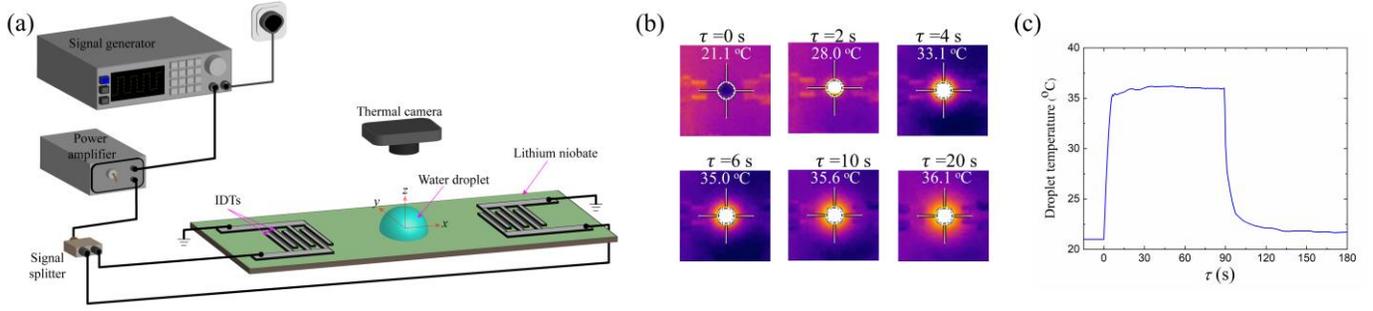

FIG. 1. Acoustothermal effect in sessile water droplets actuated by standing surface acoustic waves (SAWs). (a) Schematic of the experimental setup, wherein a standing SAW is generated via a signal generator, followed by amplification through a power amplifier and splitting by a RF splitter. A thermal camera is employed to capture the evolution of the droplet temperature. (b) Timestamps of the temperature of a $2\mu l$ sessile water droplet actuated by 32.016 MHz standing SAW, captured by a thermal camera at 0, 2, 4, 6, 10 and 20 seconds where $\tau = 0$ s marks the initiation of the standing SAW actuation. (c) The associated dynamics of the droplet temperature where $\tau = 90$ s marks when the SAW actuation is stopped.

*Theoretical framework* – Based on the inherent dynamics, the flow and temperature fields of the acoustofluidic system can be characterized by two distinct timescales: (i) fast and (ii) slow. The fast timescale is typically associated with the oscillation period of the acoustic waves and can be denoted as $t = \omega^{-1}$ where $\omega$ is the SAW angular frequency. The slow timescale ($\tau$) pertains to the thermo-hydrodynamics of the system and is generally much larger than the fast timescales ($\tau \gg t$). Regular perturbation techniques cannot be applicable since the thermo-hydrodynamic temperature rise often exceeds the acoustic temperature fields. In the context, we write any physical response $g$ associated with the acoustofluidic system as the sum of mean ($\bar{g}$) and fast oscillatory ($\tilde{g}$) components, without utilizing perturbation approach.

$$g = \bar{g}(x,y,z,\tau) + \tilde{g}(x,y,z,t) \qquad (1)$$

where acoustic fields ($\tilde{g}$) oscillate temporally with the same frequency as that of the actuation frequency and can be expressed as $\tilde{g}(x,y,z,t) = \hat{g}(x,y,z)e^{i\omega t}$. The fluid flow is governed by the continuity, momentum and energy equations for the compressible Newtonian fluids. As, for most liquids, thermal diffusivity ($D_{th}$) is of the order of $\sim 10^{-7}$ m/s and accordingly, the diffusion timescale is of the order of $\sim 10^{-1}s$ which is much larger than the fast timescale ($\sim 10^{-8}s$). Therefore, we can safely neglect the thermal diffusion term while estimating the fast oscillatory acoustic fields, yielding a direct relationship between acoustic temperature ($\tilde{T}$) and acoustic pressure ($\tilde{p}$) as $\tilde{T} = \frac{\alpha_p \bar{T}}{\bar{\rho} c_p} \tilde{p}$. The fast oscillatory acoustic fields are governed by (see sec. S3 in the SM [16])

$$i\omega \kappa_s \tilde{p} + \nabla \cdot \tilde{\mathbf{v}} = 0 \qquad (2a)$$

$$i\omega \bar{\rho} \tilde{\mathbf{v}} = -\nabla \tilde{p} + \nabla \cdot \tilde{\boldsymbol{\sigma}} \qquad (2b)$$

where $\rho$ and $\kappa_s$ denote fluid density and the isentropic compressibility, respectively. $\tilde{\boldsymbol{\sigma}} = \mu \left[ \nabla \tilde{\mathbf{v}} + (\nabla \tilde{\mathbf{v}})^T \right] + \left( \mu_b - \frac{2}{3}\mu \right) [\nabla \cdot \tilde{\mathbf{v}}] \mathbf{I}$ is the acoustic stress tensor and it depends on shear viscosity ($\mu$), bulk viscosity ($\mu_b$) and acoustic velocity ($\tilde{\mathbf{v}}$). Acoustofluidic systems also exhibit two distinct length scales: (i) one corresponding to the viscous ($\delta_{vis} = \sqrt{\frac{2\mu}{\omega\rho}}$) and thermal ($\delta_{th} = \sqrt{\frac{2D_{th}}{\omega}}$) boundary layer thicknesses (short-range acoustic fields), and (ii) the other one corresponding to the fluid bulk (long-range acoustic fields). We carried out rigorous analysis to separate both timescales and length scales to obtain the equations governing the fast oscillatory and the mean flow/ thermal fields; see sec. S4A in the SM [16]. The acoustic pressure field is governed by

$$\nabla^2 \tilde{p}_d = -k_{wave}^2 \tilde{p}_d \quad (3)$$

where subscript $d$ refers to the long-range fields; $k_{wave} = k_0 \left(1 + i\Gamma\right)^{-\frac{1}{2}}$ is the effective acoustic wavenumber in the fluid; $k_0 = \omega/c_0$ is the wave number of the unattenuated wave in the fluid; and $\Gamma = \omega(1+\beta)\mu\kappa_s$ is the damping factor associated with the viscous effect. The long-range acoustic velocity is related to the long-range acoustic pressure ($\tilde{p}_d$) as

$$\tilde{\mathbf{v}}_d = \frac{i-\Gamma}{\bar{\rho}\omega}\nabla\tilde{p}_d \quad (4)$$

The boundary conditions associated with Eq. (3) are specified as

(i) at bottom: $\mathbf{n}.\nabla\tilde{p}_d = \frac{\bar{\rho}\omega^2(\mathbf{n}.\tilde{\mathbf{u}})}{1+i\Gamma}$

(ii) at free surface: $\tilde{p}_d = 0$

where $\tilde{\mathbf{u}}$ is the bottom displacement field given by Eq. (S2a), see SM [16]. The short-range velocity can be obtained as $\tilde{\mathbf{v}}_\delta = \tilde{\mathbf{v}}_b e^{-ik_s z}$ where $\tilde{\mathbf{v}}_b = \frac{\partial \tilde{\mathbf{u}}}{\partial t}$, $\tilde{\mathbf{u}}$ is the SAW displacement field at the bottom. Furthermore, simulating the pressure acoustic model for our droplet system necessitates substantial computational resources due to the vast discrepancy between the droplet size and the SAW wavelength. To mitigate this, we used Fourier transform to compute the acoustic fields in 2D-axisymmetric computational domain; see SM, sec. S4A [16].

The mean flow and temperature fields are governed by the time averaged equations (see SM, sec. S4B [16])

$$\partial_\tau \bar{\rho} + \nabla.\left(\langle\tilde{\rho}\tilde{\mathbf{v}}\rangle + \bar{\rho}\bar{\mathbf{v}}\right) = 0 \quad (5a)$$

$$\bar{\rho}\partial_\tau\bar{\mathbf{v}} + \bar{\rho}(\bar{\mathbf{v}}.\nabla)\bar{\mathbf{v}} = -\nabla\bar{p} + \mu\nabla^2\bar{\mathbf{v}} + \mu\beta\nabla(\nabla.\bar{\mathbf{v}}) + \mathbf{F}_{ac} \quad (5b)$$

$$\bar{\rho}c_p\partial_\tau\bar{T} + \left(\bar{\rho}\bar{\mathbf{v}} + \langle\tilde{\rho}\tilde{\mathbf{v}}\rangle\right)c_p.\nabla\bar{T} - \nabla.\left(k_{th}\nabla\bar{T}\right) = \langle q_{ac}\rangle \quad (5c)$$

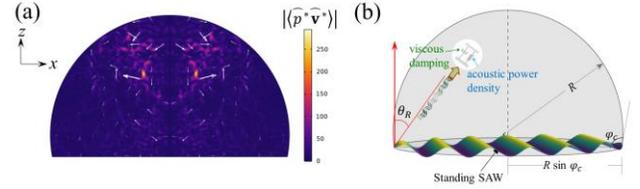

FIG. 2. Acoustic energy flow inside a 2$\mu l$ sessile droplet actuated by 32.016 MHz standing SAW. (a) Distribution of the magnitude of time-averaged acoustic energy flux shows the acoustic power flow inside the droplet at y=0 meridian plane. Two eyes are observed along the plane, corresponding to the maximum power flow regions. (b) Standing SAW mediated acoustothermal mechanism illustrating the dissipation of the acoustic energy into the droplet.

In the above $\langle\ \rangle$ denotes time-averaging of two oscillatory components, and the time-averaging is performed over the oscillation time period, $\frac{2\pi}{\omega}$ as $\langle\tilde{g}_1\tilde{g}_2\rangle = \frac{1}{2}\text{Re}\left[\tilde{g}_1^*\tilde{g}_2\right]$. Here, Re[ ] and superscript * indicate the real component and complex conjugate, respectively. The acoustic body force is given by $\mathbf{F}_{ac} = -\langle\tilde{\rho}\partial_t\tilde{\mathbf{v}}\rangle - \bar{\rho}\langle(\tilde{\mathbf{v}}.\nabla)\tilde{\mathbf{v}}\rangle$ and the acoustothermal heat source is expressed as

$$\langle q_{ac}\rangle = \nabla.\left[\langle\tilde{\mathbf{v}}.\tilde{\boldsymbol{\sigma}}\rangle - \mathbf{\Pi}_{ac}\right] \quad (6)$$

where $\mathbf{\Pi}_{ac} = \tilde{p}\tilde{\mathbf{v}}$ is the Poynting vector and it indicates time-averaged acoustic energy flux at a spatial region. Interestingly, the contribution of the term $\nabla.\langle\tilde{\mathbf{v}}.\tilde{\boldsymbol{\sigma}}\rangle$ to the acoustothermal heating is negligibly small [16,17], reducing the heat source term to

$$\langle q_{ac}\rangle = -\nabla.\langle\mathbf{\Pi}_{ac}\rangle \quad (7)$$

Above equation shows that the acoustothermal heating can be expressed as the negative divergence of the Poynting vector. We further delve into characterizing the source term and obtained the following closed form expression (see sec. S5A in the SM [16])

$$\langle q_{ac}\rangle = \Gamma\bar{\rho}\omega\langle\tilde{\mathbf{v}}.\tilde{\mathbf{v}}\rangle \quad (8)$$

This is our main result and it depicts that the acoustothermal effect, observed over slow timescale, is solely attributed to the power loss due to viscous damping. The acoustothermal heat source is proportional to the viscous damping factor, $\Gamma$, thereby revealing direct correlations with both shear and bulk viscosities. Note here, for the large systems, where fluid domain size is much larger than the acoustic wavelength, calculation of the divergence of the Poynting vector is a seemingly impossible task even with modern computational facilities. Hence, it is essential to use the closed form specified by Eq. (8) for the heating calculation. We solved Eq. (5a-b), governing the mean flow, employing the well-established limiting velocity method [18,19] where limiting velocity is prescribed at the droplet's base and the slip wall condition is prescribed at the free surface. Temperature distribution is obtained by solving Eq. (5c) along with Eqs. (5a-b) and the acoustic streaming plays a pivotal role in convective heat transport within sessile droplet. To solve Eq. (5c), at the droplet base we consider, $\bar{T} = T_0$, whereas at the free surface convective heat loss $\mathbf{n}.\left(-k_{th}\nabla\bar{T}\right) = h_{amb}\left(\bar{T} - T_0\right)$ is imposed.

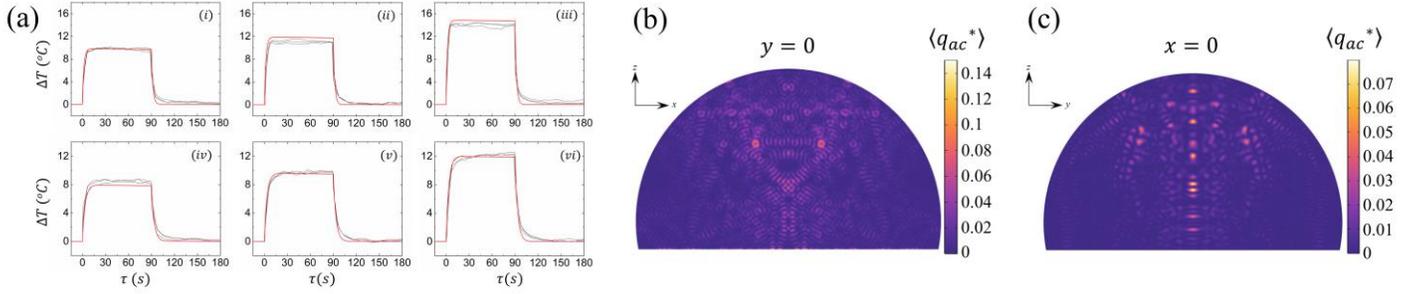

FIG. 3. Acoustothermal dynamics and heat source. (a) The dynamics of the sessile droplet temperature rise, $\Delta T$ as a function of time, $\tau$ is shown for $2\mu l$ (top row) and 5 $\mu l$ (bottom row) droplets at $P_{SAW,t}$ = 151.3 (i & iv), 183.7 (ii & v) and 232.0 (iii & vi) mW. The temperature rise data presented here is corrected considering acoustothermal effect only with no other external heat source. The black lines are experimental data whereas red lines are theoretical predictions. (b) Spatial distribution of the acoustothermal heat source is shown for $2\mu l$ droplet along meridian planes (y=0 and x=0). The maximum heat generation occurs around the two eyes correspond to the maximum Poynting vector magnitudes (see SM, sec. S6 [16]).

*Results and discussion* -In our acoustothermal experiments, the sessile droplet is excited by the standing SAW generated on a lithium niobate piezoelectric substrate (Fig. 1a), exhibiting significant temperature rise. The SAW, upon meeting, refracts as a longitudinal wave into the droplet at a Rayleigh angle $\theta_R \approx 22°$, and rapidly attenuates in a short distance, $l = \alpha^{-1} = c_0/\omega\Gamma$. The exchange of the acoustic energy occurs over fast timescale, manifesting the acoustothermal temperature rise over slow timescale (~ 10 s), as evident from the experimental temperature profiles (Fig. 1b) (**see SM video**). For a comprehensive characterization of the acoustothermal effect, we first delved into the details of the acoustic field dynamics and identified the origin of the acoustothermal heat source. The distribution of the acoustic energy flux is shown in dimensionless form in Fig. 2a where acoustic pressure and velocity fields are scaled as $\hat{p}^* = \dfrac{\hat{p}}{\bar{\rho}c_0 u_0 \omega}$, and $\tilde{\mathbf{v}}^* = \dfrac{\hat{\mathbf{v}}}{u_0 \omega}$ to effectively generalize the representation for any SAW amplitude; see SM sec. S6 [16]. The acoustic energy flux distribution is observed to be symmetrical and shows few hot spots with two prominent eyes near the droplet centroid, which results from the overlap of the reflection pathways of the longitudinal acoustic waves as they reflect multiple times on the droplet air interface and the SAW surface. The arrows in Fig. 2a indicate the direction of the power flow and it is observed to be outward in a circular fashion, leading to enhanced energy dissipation into the fluid. It is noteworthy to specify here that the significant impedance mismatch at the droplet-air interface restricts the acoustic energy from leaking through the free surface and causes large acoustic velocity as evident from Fig S4, see sec. S6 in the SM [16].

Next, we delve into the intricacies of the heating mechanism by scrutinizing the heat source term delineated in Eq. (6). Even though the term $\nabla.\left(\tilde{\mathbf{v}}.\tilde{\boldsymbol{\sigma}}\right)$ is negligibly small in contributing as the acoustothermal heat source, it can still be expanded as $\tilde{\mathbf{v}}.\left(\nabla.\tilde{\boldsymbol{\sigma}}\right) + \nabla\tilde{\mathbf{v}}:\tilde{\boldsymbol{\sigma}}$ and by employing Eq. (2b), one can reformulate Eq. (6), without time-averaging, as (see SM, sec. S5B, [16])

$$q_{ac} = \frac{\partial E_{ac}}{\partial t} + \nabla\tilde{\mathbf{v}}:\tilde{\boldsymbol{\sigma}} \qquad (9)$$

where $E_{ac} = \dfrac{\kappa_s}{2}\tilde{p}^2 + \dfrac{\bar{\rho}}{2}\tilde{\mathbf{v}}.\tilde{\mathbf{v}}$ is the acoustic energy density. The transfer of acoustic energy into the fluid is thus characterized by the rate of change in acoustic energy density and the losses associated with viscous dissipation. Necessarily, the time-averaged value of the first term on right hand side is zero, resulting the time-averaged heat source, to be only the time-averaged viscous dissipation [10] i.e.

$$\langle q_{ac}\rangle = \langle\nabla\tilde{\mathbf{v}}:\tilde{\boldsymbol{\sigma}}\rangle \qquad (10)$$

Accurately estimating the viscous dissipation term $\langle\nabla\tilde{\mathbf{v}}:\tilde{\boldsymbol{\sigma}}\rangle$ in large systems is exceedingly challenging. To overcome that, we can obtain analogous expression of the acoustothermal heat source by exploiting Eq. (7) as

$$-\nabla.\boldsymbol{\Pi}_{ac} = \frac{\partial E_{ac}}{\partial t} + \Gamma\bar{\rho}\omega\tilde{\mathbf{v}}.\tilde{\mathbf{v}} \qquad (11)$$

After time-averaging, $\left\langle \frac{\partial E_{ac}}{\partial t} \right\rangle$ becomes zero, recovering Eq. (8).

It is important to mention here that, one needs to use the closed-form expression given by Eq. (8) instead of calculating $\left\langle \nabla \tilde{\mathbf{v}} : \tilde{\boldsymbol{\sigma}} \right\rangle$ to generalize the heat source both for small and large system; see SM, sec. S7 [16]. Moreover, Eq. (11) reveals the acoustothermal mechanism and it can be effectively illustrated by the spring-dashpot system wherein the fluctuations of acoustic energy density can be conceptualized as analogous to a spring system, while the viscous damping is epitomized by the dashpot system (**Fig. 2b**).

The total heat generation is obtained by integrating Eq. (7) over the droplet volume, resulting $\left\langle Q_{ac} \right\rangle = -\int_\forall \nabla \cdot \left\langle \mathbf{\Pi}_{ac} \right\rangle d\forall$ where $\forall$ denotes the volume of the droplet. Utilizing diverging theorem, the heat source can be rewritten as $\left\langle Q_{ac} \right\rangle = -\oint_{\partial\Omega} \left\langle \mathbf{\Pi}_{ac} \right\rangle \cdot \mathbf{n} \, ds$ where $\partial\Omega$ is the closed surface of the sessile droplet. Since, the acoustic pressure, $\tilde{p}$ is zero at the free surface, the acoustothermal heat rate is the power transmitted into the droplet from the droplet bottom ($\mathbf{r} \in \partial\Omega_b$) i.e. $\left\langle Q_{ac} \right\rangle = \int_{\partial\Omega_b} \left\langle \tilde{p}\tilde{v}_z \right\rangle ds$.

In our standing SAW driven droplet experiments, we used $P_{SAW}$ = 6.1, 18.5, 36.4, 59.4, 75.6, 91.9, 116.0 mW on each IDT side. So, the total power to the droplet is given by $P_{SAW,t} = 2P_{SAW}$. Each experiment was repeated three times at each power level to ensure data reproducibility and to measure the standard deviations associated with the temperature measurements of the sessile droplets. It is should be noted that, in addition to acoustothermal effects, there are other heat generation sources such as Joule heating from the interdigital transducers (IDTs) and dielectric loss within piezoelectric substrates, responsible for the temperature rise of the droplet. Although these effects are minor, we account for the effects of these heat sources by experimentally measuring the temperature of the SAW device in absence of droplet while the SAW was activated. Henceforth, the droplet temperature rise data are to be understood as those corrected by eliminating the heating effects due to other sources, thereby quantifying the thermal profile solely due to acoustothermal effects, see SM, sec. S8 [16]. Fig. 3a illustrates the change in droplet temperature, $\Delta T$ ($= T_{\max} - T_0$), as a function of time, $\tau$ obtained experimentally and numerically, for $2\mu l$ and $5\mu l$ droplets. The numerical simulation well-captured the droplet temperature dynamics as evident from the figures. Fig. 3b presents the distribution of heat source, prescribed by Eq. (8), expressed in dimensionless form as $\left\langle q_{ac}^* \right\rangle = \Gamma \left\langle \tilde{\mathbf{v}}^* \cdot \tilde{\mathbf{v}}^* \right\rangle$ for $2\mu l$ droplet along meridian planes ($x = 0$ and $y = 0$). The spatial distribution reveals that the maximum acoustothermal heat generation is observed to occur near the two eyes associated with the maximum Poynting vector, specified earlier (FIG. 2a). It is noteworthy that, while we showed the distribution of $\Gamma \left\langle \tilde{\mathbf{v}}^* \cdot \tilde{\mathbf{v}}^* \right\rangle$, the equivalence of the heat source terms (Eq. 6 and Eq. 7) has been demonstrated both theoretically and numerically [16], as stated earlier.

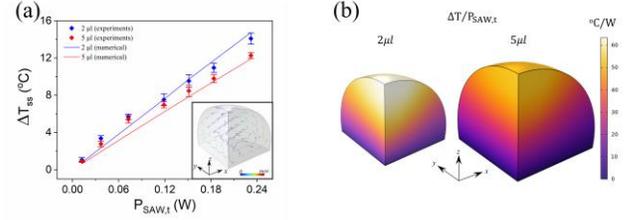

FIG. 4. Acoustothermal temperature rise. (a) Dependence of steady acoustothermal temperature rise, $\Delta T_{ss}$ on total acoustic power into the droplet, $P_{SAW,t}$ for $2\mu l$ (blue) and $5\mu l$ droplets (red). the droplets are placed on lithium-niobate-based SAW device and actuated by 32.016 MHz standing SAW. The symbols and lines indicate experimental and numerical results, respectively. the inset shows the acoustic streaming velocity vectors inside the sessile droplet where color bar indicates acoustic streaming magnitude. Due to symmetry, only one quarter of the droplet system is shown. (b) Steady temperature distribution inside one quarter of $2\mu l$ and $5\mu l$ droplets, estimated by numerical simulation. As $\Delta T$ shows linear dependence in the current operating SAW power window, the temperature distribution has been shown by scaling it with the total input SAW power.

**Fig 4a** depicts the maximum temperature rise as a function of input SAW power into the droplet of $2\mu l$ and $5\mu l$. The inset of Fig. 4a shows the arrow plot of the acoustic streaming fields in sessile droplets (due to the symmetry, only one quarter of the droplet is shown). The acoustothermal temperature rise is observed to be linearly varying with the input SAW power. It is noteworthy to mention here that the present study considered the input power up to $P_{SAW,t}$ = 232.0 mW for which the convective heat transport is not large enough to deviate the temperature rise from linearity. However, it helps in enhancing the heat transport, thereby reducing the temperature rise slightly (~ 2% drop in $\Delta T$). Experimentally, at $P_{SAW,t}$ =232 mW, the maximum temperature rise for a $2\mu l$ droplet is observed to be 14.08 $\pm$ 0.6 °C, whereas for a 5 $\mu l$ droplet, the temperature rise is 12.27 $\pm$ 0.32 °C, indicating a decrease in temperature rise with increasing droplet volume. Our theoretical estimations show a similar trend with 14.73 °C temperature rise for a $2\mu l$ droplet and 11.89 °C temperature increase for a $5\mu l$ droplet. This can be attributed to the fact that, with an increase in volume, the amount of heat required to achieve the same temperature rise increases linearly, whereas the heat generation in droplets at the same input power level does not follow a linear trend. In fact, irrespective of the power level, a $2\mu l$ droplet is able to convert 16.46% of the input power into acoustothermal heating, whereas in a $5\mu l$ droplet around 17.05% conversion of the input power is estimated. This constitutes merely a 3.6% increase in acoustic energy dissipation rate despite a 150% increase in droplet volume (and an 84.2% increase in droplet base area from where the acoustic energy is absorbed). Fig. 4b illustrates the steady state temperature distribution inside the droplet under the influence of standing SAW actuation. Due to symmetry, only one quarter of the droplet is depicted. As the temperature rise is proportional to the input SAW power, we present the temperature distribution, scaled by the SAW power for both $2\mu l$ and $5\mu l$ droplets.

With this theoretical development, we draw analogy between the acoustothermal and the electromagnetic heating. In electromagnetic theory [20], the propagation of electromagnetic wave is characterized by both electric field, $\mathbf{E}$ and magnetic field $\mathbf{H}$, where the Poynting vector $\mathbf{\Pi}_{EM} = \mathbf{E} \times \mathbf{H}$ depicts the electromagnetic energy flux. According to Poynting theorem, the generalized conservation of energy yields

$$\nabla \cdot \mathbf{\Pi}_{EM} + \frac{\partial E_{EM}}{\partial t} + \mathbf{E} \cdot \mathbf{J} = 0 \qquad (12)$$

where $E_{EM} = \frac{\varepsilon_m}{2}\mathbf{E} \cdot \mathbf{E} + \frac{\mu_m}{2}\mathbf{H} \cdot \mathbf{H}$ is the electromagnetic energy density; $\varepsilon_m$ and $\mu_m$ are the permittivity and the magnetic permeability of the medium. The conduction current density $\mathbf{J}$ is related to the electric field as $\mathbf{J} = \sigma_m \mathbf{E}$ where $\sigma_m$ is the conductivity of the medium. The net electromagnetic power flow into a spatial region is given by $-\nabla \cdot \mathbf{\Pi}_{EM}$ and it is balanced by the rate change in the electric ($\frac{\varepsilon_m}{2}\mathbf{E} \cdot \mathbf{E}$) and magnetic ($\frac{\mu_m}{2}\mathbf{H} \cdot \mathbf{H}$) energy stored along with power dissipated ($\mathbf{E} \cdot \mathbf{J}$) manifesting heat.

Analogous to electromagnetic theory, the Poynting vector in acoustic theory, $\mathbf{\Pi}_{ac} = \tilde{p}\tilde{\mathbf{v}}$ represents the acoustic energy flux. Unlike electromagnetic waves, where power flow occurs in the direction perpendicular to both the electric and the magnetic fields, the acoustic power flows coherently along the direction of the acoustic wave propagation. It is noteworthy that, Eq. 10 constitutes the Poynting theorem analogous to that of electromagnetic theory. Essentially, the electromagnetic heat source can be expressed as $\langle q_{EM} \rangle = -\langle \nabla \cdot \mathbf{\Pi}_{EM} \rangle$ similar to that of acoustic theory developed in this study. The time-averaged value of $\frac{\partial E_{EM}}{\partial t}$ is essentially zero, resulting in the electromagnetic source term as $\langle q_{EM} \rangle = \sigma_m \langle \mathbf{E} \cdot \mathbf{E} \rangle$ analogous to the acoustothermal heat source given by Eq. (8).

*Conclusion-* We investigated the origin and mechanism of acoustothermal effects in sessile water droplet system actuated by standing surface acoustic waves using experimental, theoretical, and numerical studies. We present a theory to precisely calculate the acoustic fields and the mean dynamics, without employing traditional perturbation approach. The limitation of existing theory in estimating acoustothermal heat source is discussed and a closed form heat source expression provided in this study, which is extensively validated both theoretically and numerically. An analogy with existing electromagnetic heating theory is developed to delineate the mechanism and origin of the acoustothermal effects. Analogous to electromagnetic theory, the acoustic power flows along the propagation direction of the acoustic wave and the acoustothermal heat generation is characterized by the negative divergence of the acoustic Poynting vector. The dissipation of acoustic power is shown to be dependent on the viscous damping factor, acoustic frequency and the kinetic energy density. We provided direct experimental evidence in quantifying and validating the acoustothermal dynamics in the sessile droplet system.

This work was funded by the National Science Foundation Grant No. CMI-2108795, which is gratefully acknowledged. We thank Yuqi Huang for the fabrication of the SAW device. We also thank Prof. Antoine Riaud for the assistance in developing Fourier transform based model for the acoustic field computations.

**REFERENCES**

[1] S. J. Lighthill, Acoustic streaming, J Sound Vib **61**, 391 (1978).

[2] Lord Rayleigh, On Waves Propagated along the Plane Surface of an Elastic Solid, Proceedings of the London Mathematical Society **s1-17**, 4 (1885).

[3] H. Schlichting, Über die Stabilität der Couetteströmung, Ann Phys **406**, 905 (1932).

[4] C. Eckart, Vortices and Streams Caused by Sound Waves, Physical Review **73**, 68 (1948).

[5] Prof. R. W. W. For.Mem.R.S. and A. L. Loomis, XXXVIII. The physical and biological effects of high-frequency sound-waves of great intensity, The London, Edinburgh, and Dublin Philosophical Magazine and Journal of Science **4**, 417 (1927).

[6] B. Karl Sollner, T H E MECHANISM OF T H E FORMATION O F FOGS BY ULTRASONIC WAVES, (n.d.).

[7] J. J. Markham, R. T. Beyer, and R. B. Lindsay, Absorption of Sound in Fluids, Rev Mod Phys **23**, 353 (1951).

[8] G. G. Stokes, On the Theories of the Internal Friction of Fluids in Motion, and of the Equilibrium and Motion of Elastic Solids, Mathematical and Physical Papers Vol.1 75 (2009).

[9] Samuel Temkin, *Elements of Acoustics* (John Wiley & Sons, Ltd, 1981).

[10] Philip M. Morse, *Theoretical Acoustics* (Princeton university press, 1986).

[11] R. L. Peskin, R. J. Raco, R. L. P, and R. J. Rac, Ultrasonic Atomization of Liquids, J Acoust Soc Am **35**, 1378 (1963).

[12] J. Kondoh, N. Shimizu, Y. Matsui, and S. Shiokawa, Liquid heating effects by SAW streaming on the piezoelectric substrate, IEEE Trans Ultrason Ferroelectr Freq Control **52**, 1881 (2005).

[13] R. J. Shilton et al., Rapid and Controllable Digital Microfluidic Heating by Surface Acoustic Waves, Adv Funct Mater **25**, 5895 (2015).

[14] Y. Wang et al., A rapid and controllable acoustothermal microheater using thin film surface acoustic waves, Sens Actuators A Phys **318**, 112508 (2021).

[15] Q. Y. Huang, Q. Sun, H. Hu, J. L. Han, and Y. L. Lei, Thermal effect in the process of surface acoustic wave atomization, Exp Therm Fluid Sci **120**, 110257 (2021).
[16] *See Supplemental Material*.
[17] P. K. Das, A. D. Snider, and V. R. Bhethanabotla, Acoustothermal heating in surface acoustic wave driven microchannel flow, Physics of Fluids **31**, 106106 (2019).
[18] C. Chen, S. P. Zhang, Z. Mao, N. Nama, Y. Gu, P. H. Huang, Y. Jing, X. Guo, F. Costanzo, and T. J. Huang, Three-dimensional numerical simulation and experimental investigation of boundary-driven streaming in surface acoustic wave microfluidics, Lab Chip **18**, 3645 (2018).
[19] J. Lei, P. Glynne-Jones, and M. Hill, Comparing methods for the modelling of boundary-driven streaming in acoustofluidic devices, Microfluid Nanofluidics **21**, 1 (2017).
[20] The Scattering of Light and Other Electromagnetic Radiation, The Scattering of Light and Other Electromagnetic Radiation (1969).